\documentclass[twocolumn]{aastex63}

\newcommand{\beq}{\begin{equation}}
\newcommand{\eeq}{\end{equation}}
\newcommand{\beqn}{\begin{eqnarray}}
\newcommand{\eeqn}{\end{eqnarray}}
\newcommand{\gppr}{\stackrel{>}{\scriptstyle \sim}}
\newcommand{\lppr}{\stackrel{<}{\scriptstyle \sim}}

\accepted{December 21, 2020}

\shorttitle{Turbulence and Particle Acceleration in Shearing Flows}
\shortauthors{Rieger and Duffy}

\begin{document}

\title{Turbulence and Particle Acceleration in Shearing Flows}



\correspondingauthor{Frank M. Rieger}
\email{f.rieger@uni-heidelberg.de}

\author{Frank M. Rieger}
\affiliation{ZAH, Institute of Theoretical Astrophysics, University of Heidelberg \\
Philosophenweg 12, 69120 Heidelberg, Germany}
\affiliation{Max-Planck-Institut f\"ur Kernphysik, Saupfercheckweg 1, 69117 Heidelberg, Germany}

\author{Peter Duffy}
\affiliation{School of Physics, University College Dublin, Belfield, Dublin 4, Ireland}

\nocollaboration{2}

\begin{abstract}
We explore constraints imposed by shear-driven instabilities on the acceleration of energetic 
particles in relativistic shearing flows. We show that shearing layers in large-scale AGN jets 
are likely to encompass a sizeable fraction ($\gppr 0.1$) of the jet radius, requiring seed 
injection of GeV electrons for efficient acceleration. While the diffusion process may depend 
on pre-developed turbulence if injection occurs at higher energies, electron acceleration to 
PeV and proton acceleration to EeV energies appears possible within the constraints imposed 
by jet stability. 
\end{abstract}


\section{Introduction}
The jets from Active Galactic Nuclei (AGN) exist from sub-parsec to Mpc scales \citep[see][for a recent 
review]{Blandford2019}. While apparently relativistic and uni-directional, the underlying flow is likely to 
exhibit some transverse velocity stratification already close to its origin \citep[e.g.,][]{Rieger2004}. As 
these jets then continue to propagate, interactions with the ambient medium can excite instabilities, 
induce mixing and mass loading, thereby enforcing further velocity shearing \citep[see][and references
therein]{Perucho2019}. Differences in jet speeds $V_j$ and Mach numbers $M_j$ in this regard are 
expected to contribute to the observed morphological differences of Fanaroff-Riley (FR) I and II sources, 
respectively \cite[cf.][]{Marti2019}. The growing realization of velocity shearing has in recent times 
motivated a variety of studies exploring its consequences for non-thermal particle acceleration and
emission \citep[e.g.,][]{Tavecchio2015,Rieger2016,Rieger2019b,Liang2017,Chhotray2017,Liu2017,
Kimura2018,Webb2018, Webb2019,Webb2020,Tavecchio2020}. Radio observations of AGN 
indeed provide evidence for a spine-sheath type jet structure on sub-parsec as well as on 
kiloparsc-scales \citep[e.g.,][]{Perlman1999,Laing2014,Mertens2016,Giovannini2018}.

On large scale, the survival of AGN jets is presumably largely determined by their response to the 
Kelvin-Helmholtz instability (KHI) \citep[e.g.,][]{Turland1976,Blandford1976,Birkinshaw1996,Ferrari1998,
Trussoni2008,Hardee2013}. Macroscopic KHIs can develop between two fluids in relative motion as 
present, for example, at the interface of a jet with its ambient medium. The physical cause is related to 
Bernoulli equation: If a ripple forms at the interface, the fluid has to flow faster to pass over it and thus 
exerts less pressure, allowing the ripple to grow further. This can result in mixing of the fluids and an 
effective transfer of momentum across the interface, leading to shearing, decollimation and deceleration 
of the jet. The excitation of long-wavelength KHI modes can affect the jet morphology, while short-wavelength 
modes can lead to a turbulent cascade conducive to in situ particle acceleration \citep[e.g.,][]{Ferrari1979}. 
In the present paper we are interested in whether stochastic shear particle acceleration can occur under 
conditions for which the jet stability itself is not affected.
 
Fast shear flows can facilitate energetic particle acceleration by several means \citep[see][for a recent 
review]{Rieger2019}. One of the best studied possibilities includes a stochastic Fermi-type process, in 
which particle energization occurs as a result of elastically scattering off differentially moving (magnetic) 
inhomogeneities embedded in a shearing background flow \citep[e.g.,][]{Berezhko1981,Earl1988,
Webb1989,Rieger2006,Lemoine2019}. The scattering center's speeds are then essentially defined by 
the general shear flow profile, and the efficiency determined by the flow difference sampled over one 
particle mean free path. While a strong (fast and narrow) shear layer may thus appear more favourable 
for particle acceleration, it could also be at the same time more KHI unstable. 

Given recent developments, this paper explores the regime for which both, jet stability and efficient 
shear particle acceleration are ensured. Its prime focus is on large-scale AGN jets whose 
macroscopic stability properties are accessible to a hydrodynamical description. This allows for 
a complementary view compared to a microscopic PIC approach \citep[e.g.,][]{Sironi2020}.

\section{Shear instabilities and turbulence}
Extragalactic radio jet are generally characterized by high Reynolds numbers, possibly reaching $Re
\sim 10^{28}$ \citep{Birkinshaw1996}. Hence, even if initially laminar, the flows are expected to quickly 
develop turbulent boundary layers that spread inwards and facilitate entrainment of ambient matter 
\citep[e.g.,][]{deYoung1993}. As a result of mixing, the shear layer widens, suggesting that its lateral
extent increase with distance along the jet. While the physical nature of the transverse momentum 
transport (giving rise to internal friction/viscosity) in relativistic jets is not yet understood, it presumably 
involves turbulent eddy as well as cosmic-ray viscosity \citep[i.e., associated with energetic particle 
acceleration, e.g.,][]{Earl1988,Rieger2006,Webb2019}. The boundary layer, on the other hand, is 
likely to be shaped by the nonlinear growth of unstable Kelvin-Helmholtz (KH) modes. The excitation 
of unstable, macroscopic KH modes in relativistic, kinetically-dominated cylindrical jets has been 
studied, both by analytical and numerical means, in the vortex-sheet approximation as well as for 
sheared flows \citep[e.g.,][]{Hardee1979,Birkinshaw1991,Hanasz1996,Hardee1998,Urpin2002,
Perucho2004,Perucho2005,Perucho2007,Rossi2008}. The initially exponentially growing modes 
inferred from linear stability analysis could disrupt or deform the jet if non-linear saturation does not 
occur. Numerical simulations in fact indicate that jets with high Lorentz factors and large Mach numbers 
could get stabilised by the presence of a shear layer due to various non-linear effects (e.g., shear-layer 
resonances) \citep[e.g.,][]{Perucho2007b}.

The present paper provides a first attempt to relate macroscopic, shear-driven, KH-type instabilities to 
shear particle acceleration, and explores its consequences for the layer width and the particle acceleration 
efficiency. Following \citet{Urpin2002}, we can get insights into the related KHI growth times by means 
of some basic physical considerations. For convenience, we assume a cylindrical relativistic jet with outer 
radius $R_j$ and smoothly varying (yet still relativistic) velocity shear profile $V_j(r)\,\vec{e}_z$. Due to 
the longitudinal bulk shear, the emergence of a small velocity perturbation $v_{1,r}$ in the radial direction 
will induce a perturbation in the $z$-direction, $dv_{1,z}/dt \simeq -v_{1,r}\, (dV_j/dr)$. 
Denoting the characteristic timescale of this process by $\tau$, we can write 
 \begin{equation}
  v_{1,z}\simeq - \tau\, v_{1,r}\, (dV_j/dr)\,.
\end{equation}
Following the (linearized) continuity equation, velocity perturbations are related to density perturbations
by $\frac{\partial\rho_1}{\partial t} \simeq - \rho\, (\partial/\partial z) v_{1,z}$, or $\frac{\rho_1}{\tau} \simeq
- \rho\, v_{1,z}\,k$, where $k$ is the longitudinal wave number, so that 
\begin{equation}
    \rho_1 \simeq \rho\,k \tau^2 v_{1,r} \,(dV_j/dr)\,.
\end{equation}
Since density perturbations are accompanied by pressure perturbations, $p_1=c_{sj}^2\, \rho_1$ 
(adiabatic case), we have 
\begin{equation}
p_1 \simeq  c_{sj}^2\, \rho\,k \tau^2 v_{1,r}\, (dV_j/dr)\,.
\end{equation}
Pressure forces, on the other hand, can change the radial velocity, $\Gamma_j^2 \rho\, (v_{1,r}/\tau)  
\simeq -(\partial p_1/\partial r)$, according to the (linearized) relativistic momentum equation, so that
\begin{equation}
\frac{v_{1,r}}{\tau} \simeq -\frac{1}{\rho \Gamma_j^2} \frac{\partial}{\partial r}
\left[c_{sj}^2\, \rho\, k \tau^2 v_{1,r}\, (dV_j/dr)\right]\,,
\end{equation} where $\Gamma_j(r)=1/(1-V_j(r)^2/c^2)^{1/2}$ is the bulk flow Lorentz factor.
Following \citet{Urpin2002} the radial change in the growth rate scales with velocity gradient as
$\frac{\partial}{\partial r} (1/\tau)\sim k \left(dV_j/dr\right)$, so that $(\partial \tau/\partial r) 
\sim - k \tau^2 \left(dV_j/dr\right)$. Hence, one finds that the (instability) growth rate in the galaxy 
rest frame becomes
\begin{equation}\label{growth_rate}
\frac{1}{\tau} \simeq \sqrt{\frac{\sqrt{2} c_{sj} k \,(dV_j/dr)}{\Gamma_j}} 
                     \simeq \sqrt{\frac{\sqrt{2} c_{sj} c\,k \,(d\Gamma_j/dr)}{\Gamma_j^4}}\,,
\end{equation}
where $dV_j/dr = (c^2/V_j) \,(d\Gamma_j/dr)(1/\Gamma_j^3) \simeq c \,(d\Gamma_j/dr) \times (1/\Gamma_j^3)$ 
has been used on the right hand side. This expression agrees with the more detailed derivation (eqs.~[26] and
[41]) in \citet{Urpin2002}. It implies a length scale $L$ at which a relativistic shear flow may get significantly 
destabilised of
\begin{equation}\label{disruption}
L  \simeq c\tau^* \simeq \Gamma_j^{3/2} \sqrt{\frac{c}{c_{sj}} \Delta R_j \frac{\Gamma_j}{(d\Gamma_j/dr)}} 
      \sim \Gamma^{3/2} \Delta R_j \,\sqrt{\frac{c}{c_{sj}}}\,,
\end{equation} assuming $\tau^*=3\tau$, $k \simeq 2\pi/\Delta R_j$, a shear width $\Delta R_j \leq R_j$, 
and $d\Gamma_j/dr\sim \Gamma_j/\Delta R_j$. For $\Gamma_j=10$, $\Delta R_j \sim R_j=100$ pc and 
$c_{sj}=c/10$ for example, this would yield $L \sim 10$ kpc. Equation~(\ref{growth_rate}) implies that the 
growth time $\tau$ increases with $\Gamma_j$ and $M_j=V_j/c_{sj}$, so that the shear-KHI could be 
suppressed in supersonic, highly relativistic jets. On the other hand, the instability arises faster for perturbations 
with shorter axial wavelengths, $\tau\propto 1/k^{1/2}$, suggesting that jets could possibly be made turbulent 
by shear stresses. Requiring a jet length-to-width ratio $L/R_j \geq 10$, equation~(\ref{disruption}) can be used 
to estimate the required shear width, 
\begin{equation}\label{shear_width}
\left(\frac{\Delta R_j}{R_j}\right) \gppr 0.1 \left(\frac{10}{\Gamma_j}\right)^{3/2} \left(\frac{10}{M_j}\right)^{1/2}\,.   
\end{equation} We note that this is compatible with recent findings for the large-scale jet in S5 0836+719 
\citep{Vega2019}. Equation~(\ref{shear_width}) suggests that in the absence of other stabilizing effects,
narrow ($<0.1 R_j$) shear layers are unlikely to exist around the kpc-scale jets of AGN, and this could be
or relevance for understanding the potential of shear particle acceleration.

\section{Shear particle acceleration}
Fast shear flows can facilitate a stochastic Fermi-type acceleration of energetic ($v\simeq c$) charged 
particles as these particles diffuse across the flow and sample the differences in flow speed \citep[see][for 
a recent review]{Rieger2019}. Gradual shear acceleration generally requires relativistic flow speeds 
($\Gamma_j >$ a few) to be efficient \citep{Webb2018,Rieger2019b}, and this makes an application to 
AGN jets particularly interesting.

For the relativistic shear flow profile considered above, $V_j(r) \vec{e}_z$, the characteristic 
(comoving) shear acceleration timescale can be expressed as \citep[e.g.,][]{Rieger2019}
\beq\label{tacc}
 t_{\rm acc}'(p') = \frac{c}{(4+\alpha)\, \tilde{\Gamma}_s \,\lambda'} \propto p'^{-\alpha}\,,
\eeq where $\lambda'(p') =c\,\tau_s'(p')$ is the energetic particle mean free path, $\tau_s'(p')$ is the 
(momentum-dependent) mean scattering time, assumed to follow a parameterization $\tau_s'(p') 
= \tau_0'~(p'/p_0')^{\alpha}$, with $p'=\gamma' m c$ the comoving particle momentum, and 
$\tilde{\Gamma}_s$ is the shear coefficient given by \citep{Rieger2004,Webb2018}
\beq\label{shear_coefficient}
 \tilde{\Gamma}_s = \frac{1}{15}\, \Gamma_j(r)^4 \left(\frac{\partial V_j}{\partial r} \right)^2\,.
\eeq 
In general, the acceleration timescale depends on both, the particle momentum and the 
radial coordinate.
Equation~(\ref{shear_coefficient}) indicates that a strong velocity shear (i.e., large gradients) 
will be favourable for particle acceleration. Equation~(\ref{shear_width}), on the other hand, 
constrains $\Delta R_j$ to be above some threshold for a given source. Since $\lambda$ appears 
in the denominator of equation~(\ref{tacc}), shear acceleration of electrons generally requires seed 
injection of some pre-accelerated electrons to be efficient, i.e., to proceed within the lifetime of the 
system. One can estimate the required seed energies ($\gamma_{\rm min}'$) by equating the 
acceleration timescale $t_{\rm acc}(r_0)'$ (or a multiple $\eta$ of it) with the (comoving) dynamical 
timescale, i.e. $\eta t_{\rm acc}' = t_{\rm dyn}' \equiv L/\Gamma_j V_j$, with $\eta >1$. 
As before \citep[e.g.,][]{Liu2017,Rieger2019b}, we use a quasi-linear-type parameterisation for 
the particle mean free path in the shear, i.e. 
\beq\label{lambda}
\lambda'\simeq  \xi^{-1} r_g'\left(\frac{r_g'}{\Lambda_{\rm max}}\right)^{1-q} \propto \gamma'^{2-q}\,, 
\eeq where $\xi \leq 1$ denotes the energy density ratio of turbulent versus regular magnetic field 
$B'$, $\Lambda_{\rm max}$ is the longest interacting wavelength of the turbulence, $r_g'=\gamma' 
mc^2/eB'$ is the particle Larmor radius, $\gamma'$ the comoving particle Lorentz factor, and $q$ 
is the power index of the turbulence spectrum (e.g., $q=5/3$  for Kolmogorov-type turbulence). 
For the chosen notation, $\alpha = 2-q$, i.e., $\alpha =1/3$ for a Kolmogorov-type turbulence.
We note that while eq.~(\ref{lambda}) provides a first approach, in general appropriate 
numerical simulations are needed to properly quantify the diffusive particle transport in turbulent 
shearing flows.
Assuming a linearly decreasing velocity profile, $V_j(r) = V_0 - (\Delta V_j/\Delta R_j)\,(r-r_0)$
with $r_0=0$, and $\Lambda_{\rm max} = \Delta R_j$, the above condition results in
\beq
 \gamma_{\rm min}' \geq \frac{e\,B' L}{m\,c^2}  \left[\frac{15 \xi\, \eta}{(4+\alpha) \Gamma_j^3} 
                                \left[\frac{c}{\Delta V_j}\right]^{2} \left(\frac{V_j}{c}\right)
                               \left[\frac{\Delta R_j}{L}\right]^{3-q}\right]^{\frac{1}{2-q}}
\eeq Hence, for electrons and $q=5/3$ one obtains
\beqn\label{gamma_min}
\gamma_{\rm min,e}' &\geq& 3.3\times 10^3\left(\frac{B'}{10^{-4}~\mathrm{G}}\right) 
                                     \left(\frac{5}{\Gamma_j}\right)^9 \nonumber \\
                                   &\times& \left(\frac{\Delta R_j/L}{50}\right)^4 
                                       \left(\frac{L}{5\,\mathrm{kpc}}\right)
                                         \left(\frac{\xi \eta}{1.0}\right)^3\,.
\eeqn For protons, $\gamma_{\rm min,p}' \geq \gamma_{\rm min,e}' \times (m_e/m_p)$.
Equation~(\ref{gamma_min}) indicates that efficient electron acceleration can require substantial
seed injection. It seems possible, however, that in many cases the required energetic seed 
electrons can be provided by classical, first (diffusive shock) and second-order Fermi type processes 
\citep[e.g.,][]{Liu2017}. Note that given the widths inferred in Equation~(\ref{shear_width}), 
the operation of non-gradual ($\lambda' > \Delta R_j$) shear particle acceleration has a rather high 
seed energy threshold, and this could affect inferences related to non-gradual shear acceleration 
of pick-up cosmic rays in large-scale AGN jets \citep[e.g.,][]{Kimura2018}. 

Estimating maximum achievable Lorentz factors $\gamma_{\rm max,e}'$ for electrons by 
balancing the acceleration timescale with the synchrotron loss timescale,  $t_{\rm acc}'= t_{\rm syn}' 
\equiv (9m^3c^5)/(4e^4\gamma' B'^2)$, suggests that in principle $\gamma_{\rm max,e}'\sim 10^8$ 
is possible. This makes shear acceleration an interesting candidate for understanding the origin of
the extended high-energy emission in the large-scale jets of AGN, such as in e.g. Centaurus~A 
\citep{HESS_Nat2020}.
In the case of protons, on the other hand, synchrotron losses are usually negligible, and 
achievable energies are instead limited by the condition of lateral confinement, $\lambda' \leq 
\Delta R_j$ \citep{Rieger2019b}

In general, efficient shear acceleration requires the presence of sufficient turbulence scattering 
particles across the flow. As noted before, it seems likely that shear-driven KH instabilities could 
contribute to the generation of macroscopic turbulence. Numerical simulations of mildly relativistic
(counter-propagating), transonic shear flows suggests that KHI vortices could quickly drive 
electromagnetic turbulence \citep{Zhang2009}. 
In the following we explore the case where the wavenumber of the KHI unstable modes is 
related to the energetic particle mean free path. Working in the comoving frame, we employ 
$k'= 2\pi/\lambda'$ with $k \sim k'/\Gamma_j$. We use relation~(\ref{growth_rate}) with $\tau' =
\tau/\Gamma_j$ to characterize the relevant (comoving) instability timescale $\tau'$, yielding
\beq
\tau' \simeq \left(\sqrt{2} c_{sj} (2\pi/\lambda') (dV_j/dr) \right)^{-1/2} 
        \propto \gamma'^{\frac{\alpha}{2}}\,.
\eeq The (comoving) particle Lorentz factor at which acceleration and turbulance generation 
would proceed on the same timescale is then given by
\beq
 \gamma_t' = \frac{e B'}{m c^2} \left(\frac{15}{4+\alpha}\right)^2 
                                                \left(\frac{c}{(dV_j/dr)}\right)^3 
                    \frac{ 2^{3/2}  \pi \xi^3}{\Gamma_j^{8}  M_j  (\Delta R_j)^{2}} \,.
\eeq 

\begin{figure}[t]
 \vspace*{0.2cm}
\includegraphics[width=0.48\textwidth]{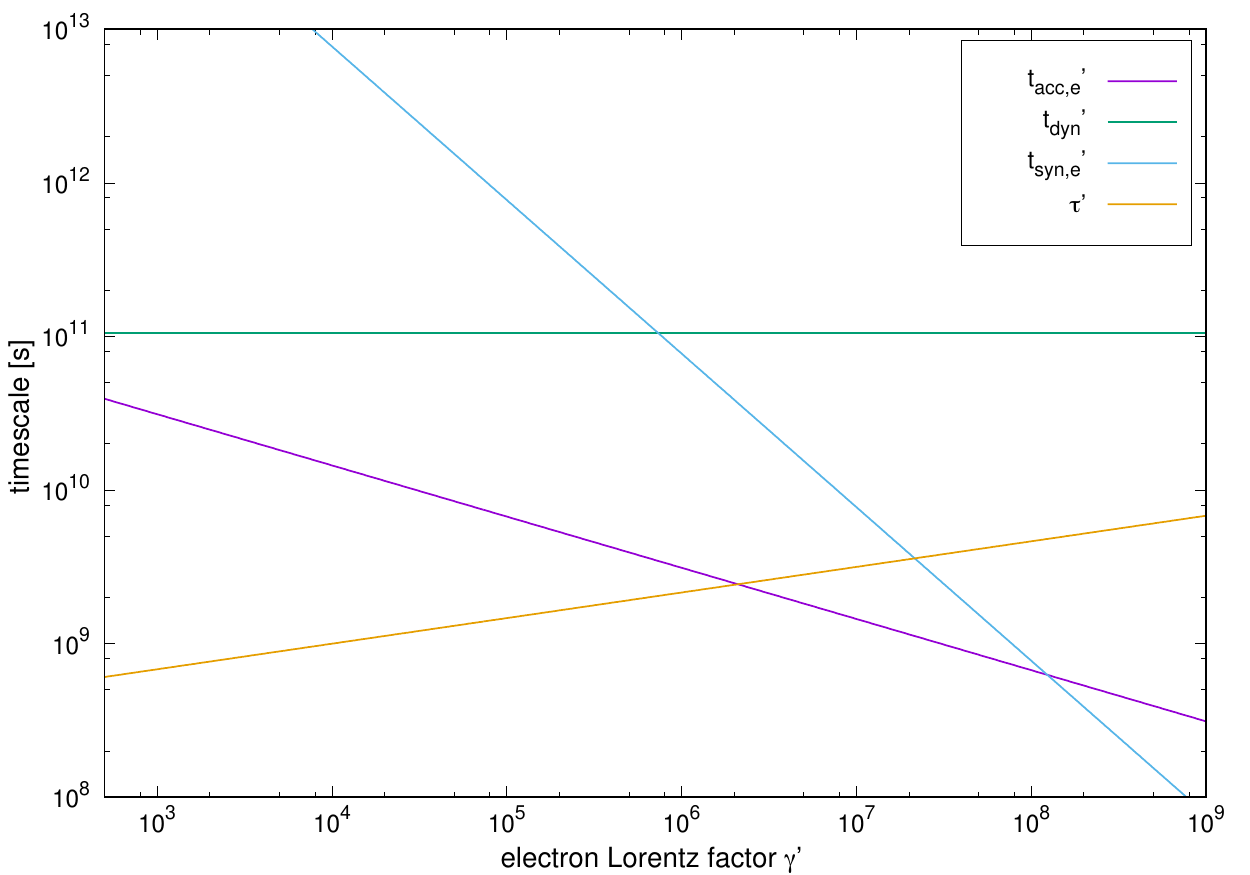}
\caption{Characteristic (comoving) timescales for electron shear acceleration. The green, 
blue, purple and yellow lines denote the dynamical timescale $t_{\rm dyn}'$, the synchrotron 
loss timescale $t_{\rm syn,e}'$, the electron acceleration timescale $t_{\rm acc,e}'$ and the 
KHI instability timescale $\tau'$, respectively. For the chosen set of parameter 
$\gamma_{\rm max,e}' \simeq 10^8$ (synchrotron-limited). Parameters used are: $L=5$ kpc, 
$\Delta R_j=0.02 L$, $B'=10^{-4}$G, $\Gamma_j=5$, $\xi=0.2$, $c_{sj}=0.05$c.}
\label{fig1}
\end{figure}

Figure~(\ref{fig1}) shows an example of the resultant characteristic timescales as applied to 
electrons, with $t_{\rm acc}' \equiv t_{\rm acc}'(r_0)$. The maximum electron energy in 
this case is synchrotron-limited to $\gamma_{\rm max,e}' \simeq 10^8$, and the ratio 
$t_{\rm dyn}'/\tau(\gamma_{\rm max,e}')' \simeq 22$. For efficient acceleration, seed electron 
injection with $\gamma_e' \gppr 10^3$ would be needed. Over a wide regime $\tau' \ll t_{\rm 
dyn}'$, suggestive of a fully developed turbulence. For $\gamma_e' \lppr \gamma_t' \simeq 
2\times 10^6$, local turbulence generation is sustained by proceeding faster than acceleration. 
For $\gamma_e' \gppr 2\times 10^6$, on the other hand, $\tau' \geq t_{\rm acc,e}'$, so that 
continuous acceleration would depend on pre-developed KHI turbulence. The situation can 
effectively be stabilized, however, if electron seed injection ($\gamma_i'$) occurs sufficiently 
below $\gamma_t'$ (e.g., at $\gamma_i' \sim \gamma_t'/5$), as $t_{\rm acc}(\gamma_i')'$ will 
set the relevant time threshold. Complementarily, interactions of the jet with stars could possibly 
provide an additional source of small-scale turbulence for the jet shear layer \cite[e.g.,][]{Perucho2020}.

\begin{figure}[t]
\vspace*{0.2cm}
\includegraphics[width=0.48\textwidth]{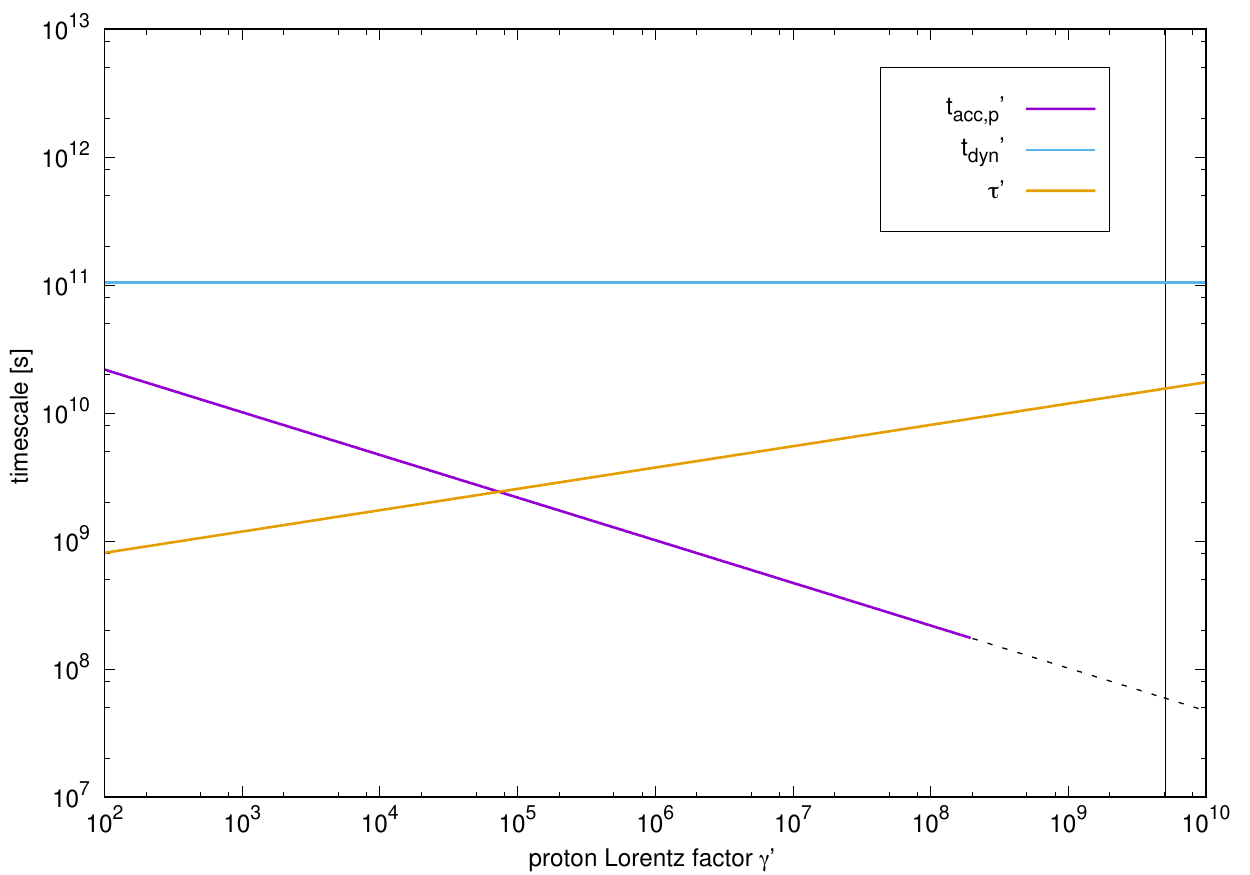}
\caption{Characteristic (comoving) timescales for proton shear acceleration. The green, 
purple and blue lines denote the dynamical timescale $t_{\rm dyn}'$, the proton acceleration 
timescale $t_{\rm acc,p}'$ and the KHI instability timescale $\tau'$, respectively. The vertical
black line denotes the confinement limit $\gamma_{\rm max,p}' =5\times 10^9$. Parameters 
used are: $L=5$ kpc, $\Delta R_j=0.02 L$, $B'=10^{-4}$G, $\Gamma_j=5$, $\xi=0.8$, $c_{sj}
=0.05$c.}\label{fig2}
\end{figure}

Figure~(\ref{fig2}) shows an illustration of the resultant characteristic timescales for protons, 
with $t_{\rm acc}' \equiv t_{\rm acc}'(r_0)$. The dotted line characterises the regime 
where, formally, the diffusion approximation employed in gradual shear becomes violated, 
and transition to non-gradual shear is expected to occur \citep[cf.][]{Rieger2019}.
The maximum proton energy in this case is expected to be limited to $\gamma_{\rm max,p}' 
=5\times 10^9$, and the ratio $t_{\rm dyn}'/\tau(\gamma_{\rm max,p}')' \simeq 7$. Since we 
anticipate the layer stability ($\Delta R_j$) to be only disrupted after several e-folding times 
($\tau'$), such (comoving) energies $\sim \gamma_{\rm max,p}'$ may still be achievable.
For seed injection $\gamma_p' \gppr \gamma_t' \simeq 7\times 10^4$, acceleration would 
depend on pre-developed KHI turbulence or other turbulence driving mechanisms. 
In principle, significant turbulence damping and jet deceleration might become possible for 
low jet powers or sufficiently high cosmic-ray seed densities. 
For example, if a seed density comparable to that of the cosmic rays in our galaxy ($\sim 3
\times 10^{-15}$ cm$^{-3}$) would be picked up at TeV energies, then this may become 
apparent for jets with powers $L_j \lppr 10^{43}$ erg/s. 

\section{Conclusion}
The foregoing analysis suggests that relativistic AGN jets can get stabilized by the presence 
of an extended ($\gppr 0.1 R_j$) shear layer. This inference is based on an simplified linear 
model of KH-driven shear instabilities in a steady, kinetically-dominated flow. In general, KH
instabilities may not necessarily destroy the jet configuration, as the system could instead 
saturate leading to another stable configuration, and inclusion of a parallel magnetic field 
could contribute to jet stability \citep[e.g.,][]{Hamlin2013}. In addition, the jet stability 
properties are likely to be affected by the characteristics of the confining medium.
While a more complex application is thus desirable and aimed for in further work, we 
note that our current estimates are compatible with recent experimental findings 
\citep{Vega2019}. As found here, efficient particle acceleration in extended shear layers 
usually requires seed injection of energetic GeV) electrons. Shear acceleration of electrons
(synchrotron-limited) to multi-TeV energies and protons (confinement-limited) to EeV energies 
seems then feasible in the kpc-scale jets of AGN \citep{Rieger2019b}. As shown here, 
shear-driven instabilities could in fact contribute to turbulence generation and thereby facilitate 
particle acceleration. Relating the particle mean free path to the wavelength of unstable modes 
indicates that jet stability could be ensured and suitable turbulence be produced. This suggests 
that shear particles acceleration could represent a viable mechanism for the energization of 
charged particles in the relativistic jets of AGN. The above considerations are expected 
to inform our understanding towards a desirable, multi-scale treatment 
\citep[e.g.,][]{Marcowidth2020} of particle acceleration in relativistic shearing flows.

\acknowledgments
We appreciate stimulating discussions with Tony Bell out of which this study grew, and 
are very grateful to Manel Perucho for comments on an earlier version of the manuscript. 
We thank the referee for useful comments.
FMR acknowledges funding by a DFG Heisenberg Fellowship under RI 1187/6-1.

\bibliography{references}{}
\bibliographystyle{aasjournal}

\end{document}